\documentstyle[12pt,epsfig]{article}
\def\lapprox{\lower .7ex\hbox{$\;\stackrel{\textstyle <}{\sim}\;$}}
\def\gapprox{\lower .7ex\hbox{$\;\stackrel{\textstyle >}{\sim}\;$}}

\def\d{{\rm d}}

\def\xp{x_{I\! P}}

\begin{document}
\begin{titlepage}
\vspace*{-1cm}
\begin{flushright}
DESY--99--065 \\
TTP99--26\\
June 1999 \\
\end{flushright}                                
\vskip 3.5cm
\begin{center}
{\Large\bf Spin Dependence of Deep Inelastic Diffraction}
\vskip 1.cm
{\large  J.~Bartels$^a$, T.~Gehrmann$^b$ and M.G.~Ryskin$^c$} 
\vskip .7cm
{\it $^a$ II.~Institut f\"ur Theoretische Physik, Universit\"at Hamburg,
D-22761 Hamburg, Germany}
\vskip .4cm
{\it $^b$ Institut f\"ur Theoretische Teilchenphysik,
Universit\"at Karlsruhe, D-76128 Karlsruhe, Germany}
\vskip .4cm
{\it $^c$ Petersburg Nuclear Physics Institute, 188350 Gatchina, 
St.~Petersburg, Russia}
\end{center}
\vskip 2.6cm

\begin{abstract}
Using a perturbative model for diffractive interactions, we derive 
an expression for 
the polarized diffractive structure function $g_1^D$ in the high energy 
limit. This structure function is given by the
interference of diffractive amplitudes with polarized and unpolarized
exchanges. For the polarized exchange we consider both
two-gluon and quark-antiquark amplitudes. 
The polarized diffractive amplitude receives sizable
contributions from non-strongly ordered regions in phase space,
resulting in a double logarithmic enhancement at small $x$. The 
resummation of these double logarithmic terms is outlined. We also discuss the
transition from our perturbative expression to the 
 nonperturbative region.
A first numerical estimate indicates that the perturbative contribution
to the spin asymmetry is substantially larger than the nonperturbative one.
\end{abstract}
\vfill
\end{titlepage}                                                                
\newpage                                                                       

\section{Introduction}

The occurrence of large rapidity gaps between current jet direction and 
proton remnant direction in electron-proton collisions at HERA
represents
one of the most puzzling phenomena in the physics of deep inelastic 
scattering. The first observation of these diffractive 
deep inelastic scattering events 
has triggered much theoretical effort towards gaining an understanding of this 
phenomenon. Although much progress both in the theoretical 
description and in the experimental study  of diffraction 
in deep inelastic scattering has been made, it is still fair to 
say that this phenomenon is not 
unambiguously understood at present, since it contains both perturbative and
nonperturbative components.

To summarize our present understanding of (unpolarized) diffraction in
deep inelastic scattering, the diffractive final states
can be attributed either to nonperturbative soft exchanges
or to hard exchanges, represented by the (slightly nonforward)
gluon structure function. Examples of the former class are the aligned jet 
configuration of diffractively produced quark-antiquark pairs 
or the analogous quark-antiquark
gluon configuration where the gluon is rather soft (in particular, has a small
transverse momentum). For the latter class of final states, we mention the
diffractive production of longitudinal vector particles or quark-antiquark 
jets with large transverse momenta. One of the most prominent signatures
of these distinct classes is the energy dependence: for the nonperturbative
part one expects to see the energy dependence of the soft Pomeron,
whereas the perturbative part should be characterized by the stronger 
increase with energy as observed in the gluon structure function at
small $x$, as determined by perturbative QCD. 
The diffractive contribution to deep inelastic structure functions
can not be easily attributed to one of both classes. 
Experimental data on it lie indeed  between the two extremes, thus 
indicating the existence of sizable
contributions from both  hard and soft exchanges.

The currently discussed option of operating the HERA collider with a 
polarized proton beam naturally leads to the expectation that, again,
DIS diffraction will be an important phenomenon, and hard and soft physics 
will compete with each other. In particular, one might expect that, as in the
unpolarized case, a substantial part of the diffractive cross section may be 
calculable within perturbative QCD.

Recently~\cite{mana} a detailed study of the non-perturbative contributions 
to spin asymmetries in deep inelastic diffraction has lead to the conclusion
that these are very small. The investigation has been carried out in the 
framework of Regge theory: the amplitude for diffraction is described by 
Pomeron pole exchange and has a (small) spin-dependent component. It was found 
that this amplitude alone yields a vanishing spin asymmetry.
Non-zero asymmetries can be obtained only if, in addition to the simple 
Pomeron pole, also secondary Reggeon exchanges 
($\rho$, $\omega$, $f$, $A_2$) as well as multi-Pomeron and Pomeron-Reggeon 
cuts are taken into account. Contributions due to multi-Pomeron cuts turn out 
to be negligible small. The 
dominant contribution to diffractive spin asymmetries arises from the 
interference of the amplitudes for Pomeron-Reggeon and 
single Reggeon exchange. As a result, the polarized cross section is 
suppressed, in comparison with the unpolarized cross section, by one inverse 
power of the collision energy. Another nonzero contribution is due to the 
exchange of unnatural parity ($\pi$, $a_1$) which, again, is suppressed by
one inverse power of energy. The numerical study of all these contributions
shows that, in addition to the energy-dependent suppression, they also come 
with very small coefficients. In total, the resulting 
nonperturbative contribution to the spin asymmetry does not exceed 
$10^{-4}$~\cite{mana}.

This raises the interesting possibility that, unlike in the unpolarized case
where the nonperturbative contributions are not small, polarized diffraction
may be dominated by the perturbative component. In perturbative QCD the
diffractive exchange is modelled by the (slightly nonforward) unpolarized 
gluon structure 
function $g(x,\mu^2)$, whereas for the polarized part 
one uses the polarized gluon density $\Delta g(x,\mu^2)$ and 
quark density $\Delta q(x,\mu^2)$. In this
framework the spin asymmetry is then described by the interference of the 
unpolarized and polarized diffractive amplitudes. Since at small $x$ the
polarized quark and gluon densities
-- apart from logarithmic corrections -- are known to 
be suppressed by one power of energy
compared to the unpolarized gluon density, the asymmetry, when calculated
perturbatively, has the same energy dependence as the nonperturbative 
contributions obtained from Regge theory~\cite{mana}. An important
enhancement of the perturbative contributions, however, is due to the
logarithmic corrections which lead to a strong rise of the polarized 
gluon distribution at small $x$. 
In~\cite{ber} it has been shown that, unlike the unpolarized case,
the polarized structure functions have double logarithms in $1/x$,
which result in a substantial enhancement of the polarized distributions. 
Applied to diffractive phenomena, this enhancement implies that the ratio of 
perturbative to non-perturbative contributions could be much more 
favourable in the polarized case than in the unpolarized case. 

In the present paper, we investigate the perturbative contribution to 
spin asymmetries in deep inelastic scattering by computing the 
longitudinal spin-spin asymmetry for the diffractive production of light 
quark-antiquark pairs. For the polarized exchange we include both two-gluon
and quark-antiquark exchange.
Basic features of the spin dependent cross section 
are first elaborated for a quark target. 
We demonstrate that the spin dependent 
part of the amplitude receives an essential contribution from a non-strongly 
ordered region, where the transverse momenta of the scattered quarks are 
lower than transverse momenta occurring in the exchange system. These 
contributions  give rise to double logarithmic terms. Including the
double logarithmic results for 
the polarized amplitudes from~\cite{ber} we obtain a compact
expression for the perturbative contribution to the asymmetry. Precise
numerical predictions of this asymmetry would require knowledge on the
behaviour of polarized quark and gluon distributions at small $x$, which 
are only poorly determined at present. A rough estimation of the
perturbative contribution  does however indicate the resulting
asymmetry to be substantially larger than the nonperturbative
contributions computed in~\cite{mana}.

\section{Kinematics and basic formulae}

Cross sections for diffractive DIS are 
defined in close analogy to the cross sections in inclusive DIS.
 A summary of formulae for the inclusive case can for 
example be found in~\cite{badelek}. Since we want to define a consistent
framework for spin asymmetries in diffractive DIS, let us briefly 
review the conventionally used definitions for polarized cross sections 
and asymmetries. 

In lepton-nucleon scattering with longitudinally polarized 
lepton beam and nucleon target, one can choose four different combinations 
of beam and target polarizations. If only parity conserving interactions 
are relevant,
these can be expressed in 
terms of two independent cross sections:
\begin{equation}
\d \Delta \sigma \equiv \left( \d 
\sigma^{\stackrel{\rightarrow}{\Leftarrow}}
- \d \sigma^{\stackrel{\rightarrow}{\Rightarrow}}\right)\;, \qquad
 \d \bar{\sigma} \equiv \frac{1}{2} \left( \d 
\sigma^{\stackrel{\rightarrow}{\Leftarrow}}
+ \d \sigma^{\stackrel{\rightarrow}{\Rightarrow}}\right)\; ,
\end{equation}
where the arrows denote the spin directions of beam and target. 
Using these, one obtains the longitudinal spin asymmetry
\begin{equation}
A_{L} = \frac{\Delta \sigma}{2\bar \sigma}\; . 
\end{equation}
The above cross sections define the unpolarized and polarized structure 
functions as described in~\cite{badelek}.

Processes in deep inelastic scattering can be conveniently described as
product of a virtual photon flux factor and a reduced photon-proton 
cross section. Unpolarized 
and polarized cross sections at small $x$ read then:
\begin{eqnarray}
\frac{\d \sigma}{\d x \d Q^2} & = & \frac{\alpha_{em}}{\pi x Q^2}
\left(1-y + \frac{y^2}{2}\right) \; \sigma_T^{\gamma^* p} 
\; , \\
\frac{\d \Delta \sigma}{\d x \d Q^2} & = & \frac{\alpha_{em}}{\pi x Q^2}
\left(y - \frac{y^2}{2}\right) \; \Delta \sigma^{\gamma^* p}\; .
\end{eqnarray}
In the expression for the unpolarized cross section, we have restricted 
ourselves to the (dominant) contribution 
$\sigma_T^{\gamma^* p} $ from the average over the two transverse 
photon polarizations. The full unpolarized deep inelastic cross section 
is obtained by adding also the contribution from longitudinal photon 
polarization. The polarized photon-proton cross section 
$\Delta \sigma^{\gamma^* p}$ corresponds to the difference of the two 
transverse polarizations 
of the virtual photon.

The kinematics of the diffractive reaction 
\begin{displaymath}
e(l) \;  p (p)\;  \longrightarrow\;  e(l')\; p(p')\; X (p_X)
\end{displaymath}
are described by the invariants
\begin{displaymath}
s = (l+p)^2 \;,\qquad
Q^2 = -(l-l')^2\; , \qquad
\hat s = (l-l'+p)^2 \; , \qquad
M^2 = p_X^2\; , \qquad
t = (p-p')^2 \; .
\end{displaymath}
From these, we define the commonly used dimensionless parameters
\begin{displaymath}
x = Q^2/\hat s\; , \qquad 
y = \hat s/s\; , \qquad
\beta = Q^2/(Q^2+M^2) \; , \qquad
\xp = x/\beta\; . 
\end{displaymath}
Using these variables, one can express the cross section for 
unpolarized diffraction in terms of the diffractive deep inelastic 
structure function $F_T^D$:
\begin{equation}
\frac{\d \sigma}{\d \beta \d Q^2 \d \xp \d t} = \frac{4 \pi \alpha_{em}^2}
{\beta Q^4} \; \left( 1- y + \frac{y^2}{2} \right) \; F_T^{D} (\beta, Q^2,
\xp,t) \; .
\label{eq:f2d}
\end{equation}
Analogously, we define the polarized diffractive structure function 
$g_1^D$ by
\begin{equation}
 \frac{\d \Delta \sigma}{\d \beta \d Q^2 \d \xp \d t} = \frac{16 \pi 
\alpha_{em}^2}
{Q^4} \; \left( y- \frac{y^2}{2} \right) \; g_1^{D} (\beta, Q^2,
\xp,t) \; .
\label{eq:g1d}
\end{equation}

In the following section, we will derive $g_1^D$ in the framework of 
a perturbative model for diffractive reactions. Within the same model, 
$F_T^{D}$ was derived in various places in the literature~\cite{ftd,wusthoff}.
To illustrate similarities and differences between unpolarized and 
polarized calculation, we will discuss $g_1^D$ and $F_T^{D}$ in parallel below.

\section{Calculation of $g_1^D$}
\label{sec:calc}

In a perturbative framework, hard diffraction is described by the 
exchange of two partons in a color singlet state between the target 
hadron and a partonic fluctuation of the incoming virtual photon, which is 
converted into a diffractive system of mass $M$. The corresponding partonic 
reaction reads:
\begin{displaymath}
\gamma^*(q)\;  p(p) \;  \longrightarrow q(k_q) \bar q(k_{\bar q})\; p(p')\; .
\end{displaymath}
We use a Sudakov-parametrization with an auxiliary vector 
$q' = q + (x/\hat s) p$ to describe the partonic momenta 
in this reaction
\begin{eqnarray}
k_q & = & \alpha q' + \frac{k_t^2}{\alpha \hat s} + \vec{k_t}\; , \\
k_{\bar q} & = & (1-\alpha) q' + \frac{k_t^2}{(1-\alpha) \hat s} 
- \vec{k_t}\;, \\
l_n & = & \alpha_n q' + \beta_n p + \vec{l_{t,n}} \; .
\end{eqnarray}
The integral over the exchange loop momentum reads in these parameters
\begin{equation}
\d^4 l_n = \frac{\hat s}{2} \; \d \alpha_n \d \beta_n \d^2 l_{t,n}\; .
\end{equation}

\begin{figure}[t]
\begin{center}
~ \epsfig{file=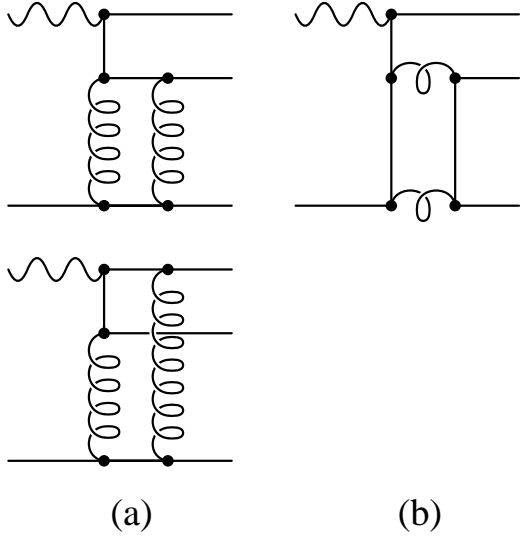,width=7cm}
\end{center}
\caption{Partonic processes contributing to diffractive $q\bar q$
  production off a quark target: (a) $t$-channel gluon exchange, (b)
  $t$-channel quark exchange.}
\label{fig:feyn}
\end{figure}

The corresponding 
Feynman diagrams for 
the two gluon exchange and quark/antiquark exchange 
amplitudes are 
depicted in Fig.~\ref{fig:feyn}. 
The cross section for this process 
reads
\begin{equation}
\frac{\d \Delta \sigma^{\gamma^*p}}{\d t \d M^2} = \frac{1}{(4\pi)^4} \; 
\frac{1}{\hat s^2}\; [\Delta] \big|{\cal M}\big|^2 
\frac{d^2 k_t \; \d \alpha}{ \alpha
(1-\alpha)}\; \delta \left( M^2 - \frac{k_t^2}{\alpha (1-\alpha)}\right) \; . 
\label{eq:diffX}
\end{equation}
Comparison with Eqs.~(\ref{eq:f2d}),(\ref{eq:g1d}) yields the diffractive 
structure functions
\begin{eqnarray}
F_T^D(\beta, Q^2,
\xp,t) & = & \frac{Q^4}{4\pi^2 x \alpha_{em}}\, \frac{\d \sigma_T^{\gamma^* p}}
{\d t \d M^2}\; ,\label{eq:ftdqcd} \\
\beta g_1^D(\beta, Q^2,
\xp,t) & = & \frac{Q^4}{16\pi^2 x \alpha_{em}}\, \frac{\d \Delta 
\sigma^{\gamma^* p}}
{\d t \d M^2} \; \label{eq:g1dqcd} .
\end{eqnarray}
In the following, we shall only consider the 
situation of forward diffraction, i.e.~$t=0$. Moreover, we work in the 
high energy (small-$x$) limit $\hat s \to \infty$, thus keeping only 
the leading power in $\hat s$.   

The diffractive matrix element 
${\cal M}$ can be decomposed into a perturbatively 
calculable part for the scattering of the two parton system 
with the virtual photon and an unintegrated structure function 
$\phi(l_t^2,\xp)$ parametrizing the probability of finding 
a parton pair with transverse momentum $\l_t$ and $-l_t$ inside the proton. 
This unintegrated structure function is a non-perturbative object, and 
a priori unknown. Its integral over $l_t^2$ generates, at leading 
logarithmic accuracy, the well-known parton distribution functions:
\begin{eqnarray}
\int^{\mu^2} \d l_t^2 \; \phi_g(l_t^2,\xp) & = & \xp g(\xp,\mu^2)\; 
,\label{eq:gllog} \\
\int^{\mu^2} \d l_t^2 \; \phi_{\Delta g}(l_t^2,\xp) & = & \xp \Delta 
g(\xp,\mu^2)\;,
\label{eq:dgllog} \\
\int^{\mu^2} \d l_t^2 \; \phi_{\Delta q}(l_t^2,\xp) & = & \xp \Delta
q (\xp,\mu^2)\; , \\
\int^{\mu^2} \d l_t^2 \; \phi_{\Delta \bar q}(l_t^2,\xp) & = & \xp \Delta
\bar q (\xp,\mu^2)\; . \label{eq:dqbllog}
\end{eqnarray}
To investigate the structure of the polarized diffractive matrix element in 
more detail, it turns out to be convenient to consider first diffraction 
off a quark target, where the full matrix element can be calculated 
perturbatively. In this case, the unintegrated structure functions reduce to
simple perturbative splitting functions. The relation between quark and 
proton matrix elements is given by the replacement
\begin{equation}
  \frac{\alpha_s}{2\pi} \, 
\lim_{\xp \to 0} \left(\xp P_{iq} (\xp)\right) \, \int \frac{\d l_t^2}{l_t^2}
\quad \longrightarrow \quad \int \d l_t^2\; \phi_i (l_t^2, \xp).
\label{eq:replace}
\end{equation}

In unpolarized diffraction, only the two gluon exchange amplitude,
Fig.~\ref{fig:feyn}.a, yields a dominant contribution, the quark-antiquark 
amplitude, Fig.~\ref{fig:feyn}.b, is suppressed by one power of $x$. 
In the polarized case, both amplitudes contribute with the same 
powers in $x$, as be shall demonstrate by explicit calculation below. 

To study the helicity structure of the diffractive exchange, let us 
first consider the square of the two gluon exchange amplitude, 
Fig.~\ref{fig:feyn}.a. Taking the target trace for an unpolarized/polarized
quark target, 
one finds in the high energy limit
\begin{eqnarray}
{\rm Tr} \left(\not \! p \gamma^{\mu_2} (\not \! p -\not \! l_2) 
\gamma^{\nu_2} \not \! p'
\gamma^{\nu_1} ( \not \! p - \not \! l_1) \gamma^{\mu_1}\right)
& = & 32 \; p^{\mu_1} p^{\nu_1} p^{\mu_2} p^{\nu_2}\; ,\label{eq:utrace}\\
{\rm Tr} \left(\gamma_5
\not \! p \gamma^{\mu_2} (\not \! p -\not \! l_2) 
\gamma^{\nu_2} \not \! p'
\gamma^{\nu_1} ( \not \! p - \not \! l_1) \gamma^{\mu_1}\right)
& = & 16 i \; \big(
p^{\mu_1} p^{\nu_1} \epsilon^{\mu_2 \nu_2
\rho \sigma} p_{\rho} \l_{2,\sigma} 
\nonumber \\ & & 
\hspace{0.5cm}
- p^{\mu_2} p^{\nu_2} \epsilon^{\mu_1 \nu_1
\rho \sigma} p_{\rho} \l_{1,\sigma} \big)\; . \label{eq:ptrace}
\end{eqnarray}
Contraction of these formulae with the $t$-channel gluon propagators
\begin{displaymath}
-g^{\alpha \alpha'} = -g_t^{\alpha \alpha'} + \frac{p^{\alpha} q'^{\alpha'} + 
p^{\alpha'} q'^{\alpha} }{p\cdot q'}
\end{displaymath}
shows that all exchanged gluons in the unpolarized $|{\cal M}|^2$ 
carry longitudinal polarization, in the polarized $\Delta |{\cal M}|^2$ only 
one of the gluons is transversely polarized. The factorization of the 
polarized target trace into  the structures $p^{\mu_i} p^{\nu_i}$ and 
$\epsilon^{\mu_j \nu_j
\rho \sigma} p_{\rho} \l_{j,\sigma}$ illustrates moreover that the 
polarized  $\Delta |{\cal M}|^2$ is the product of an amplitude
with  polarized gluon exchange with the well known unpolarized amplitude,
as first pointed out in~\cite{ryskin}. 

The matrix elements for the process 
 $\gamma^* q \to q q \bar q$ can be 
written in a form which makes this factorization explicit:
\begin{eqnarray}
|{\cal M}|^2 & = & \sum_q R_q\; \hat s\;
 \alpha (1-\alpha) \left(\alpha^2 +(1-\alpha)^2\right) \nonumber \\
& & \hspace{1cm} \times \left\{  \frac{2 k_t^i}{\bar Q^2 + k_t^2} 
- \frac{\left(k_t + l_{t,1}\right)^i}{ \bar Q^2 + (k_t+l_{t,1})^2}
- \frac{\left(k_t - l_{t,1}\right)^i}{ \bar Q^2 + (k_t-l_{t,1})^2}\right\}
\nonumber \\
& & \hspace{1cm} \times \left\{  \frac{2 k_t^i}{\bar Q^2 + k_t^2} 
- \frac{\left(k_t + l_{t,2}\right)^i}{ \bar Q^2 + (k_t+l_{t,2})^2}
- \frac{\left(k_t - l_{t,2}\right)^i}{ \bar Q^2 + (k_t-l_{t,2})^2}\right\}
\; ,\label{eq:m2} \\
\Delta |{\cal M}|^2_g & = & -\sum_q 2R_q\; \Bigg\{ (1-2\alpha)^2\; l_{t,1}^2\;
\left( \frac{2 k_t^i}{\bar Q^2 + k_t^2} 
- \frac{\left(k_t + l_{t,1}\right)^i}{ \bar Q^2 + (k_t+l_{t,1})^2}
- \frac{\left(k_t - l_{t,1}\right)^i}{ \bar Q^2 + (k_t-l_{t,1})^2}\right) 
\nonumber \\
&& \hspace{0.6cm} 
+\; 2 \left(\alpha^2 +(1-\alpha)^2\right) \nonumber \\
&& \hspace{1.2cm} \times
\Bigg(\; \frac{1}{\bar Q^2 + (k_t+l_{t,1})^2} \left( l_{t,1}\cdot (l_{t,1} + 
k_t ) k_t^i - k_t\cdot (l_{t,1} +k_t) l_{t,1}^i \right)\nonumber \\
&& \hspace{1.36cm}
+\; \frac{1}{\bar Q^2 + (k_t-l_{t,1})^2} \left( l_{t,1}\cdot (l_{t,1} - 
k_t ) k_t^i - k_t\cdot (l_{t,1} -k_t) l_{t,1}^i \right) \Bigg) \Bigg\} 
\nonumber \\
& & \hspace{1cm} \times \left\{  \frac{2 k_t^i}{\bar Q^2 + k_t^2} 
- \frac{\left(k_t + l_{t,2}\right)^i}{ \bar Q^2 + (k_t+l_{t,2})^2}
- \frac{\left(k_t - l_{t,2}\right)^i}{ \bar Q^2 + (k_t-l_{t,2})^2}\right\}
\; ,\label{eq:dm2g}\\
\Delta |{\cal M}|^2_q & = & -2R_q\; \frac{C_F}{T_F} \; \alpha (1-2\alpha) \;
l_{t,1}^2\;  \frac{k_t^i}{\bar Q^2 + k_t^2}\nonumber \\
 & & \hspace{1cm} \times \left\{  \frac{2 k_t^i}{\bar Q^2 + k_t^2} 
- \frac{\left(k_t + l_{t,2}\right)^i}{ \bar Q^2 + (k_t+l_{t,2})^2}
- \frac{\left(k_t - l_{t,2}\right)^i}{ \bar Q^2 + (k_t-l_{t,2})^2}\right\}
\; ,\label{eq:dm2q}\\
\Delta |{\cal M}|^2_{\bar q} & = & -2R_q\; \frac{C_F}{T_F} 
\; (1-\alpha) (1-2\alpha) \;
l_{t,1}^2\;  \frac{k_t^i}{\bar Q^2 + k_t^2}\nonumber \\
 & & \hspace{1cm} \times \left\{  \frac{2 k_t^i}{\bar Q^2 + k_t^2} 
- \frac{\left(k_t + l_{t,2}\right)^i}{ \bar Q^2 + (k_t+l_{t,2})^2}
- \frac{\left(k_t - l_{t,2}\right)^i}{ \bar Q^2 + (k_t-l_{t,2})^2}\right\}\; .
\label{eq:dm2qb}
\end{eqnarray}
The common factor from couplings, colour structure and from the 
integration over both  $t$-channel loops reads:
\begin{equation}
R_q = 512 \pi  \; \alpha_s^4 \; \alpha_{em} e_q^2 \; \frac{T_F^2C_F^2}{N_c}
 \; \hat s 
\; \frac{\d^2 l_{t,1}}{l_{t,1}^4} \;
\frac{\d^2 l_{t,2}}{l_{t,2}^4}\; , 
\end{equation}
where $T_F=1/2$ and 
the $l_{t,i}^2$-integrations are bound by $l_{t,i}^2<$ min$(\alpha,
1-\alpha) \hat s$ due to the requirement of an on-shell cut across the 
exchange loop.  
The subscript $g,q,\bar q$
on the polarized $\Delta |{\cal} M|^2$ denotes the 
nature of the $t$-channel exchange in the polarized amplitude.
The antiquark exchange amplitude is obtained 
by considering diffraction off an antiquark target. Note that the replacement 
(\ref{eq:replace}) acquires a negative sign if applied to a polarized 
antiquark target.

The polarized matrix elements are suppressed by one power of $\hat s$ with 
respect to the unpolarized matrix element. This suppression can be understood
from the structure of the target trace, eqs.~(\ref{eq:utrace}),
(\ref{eq:ptrace}). Contraction with the photon trace yields one power 
of $\hat s$ for every occurrence of $p^{\mu}$, while $\l_n^{\mu}$ can 
only yield terms of the type $l_{t,n}^2$ and $l_{t,n}\cdot k_t$. 
This one extra power of $\hat{s}$ is compensated by
$1/\xp$ appearing in the unpolarized quark-to-gluon splitting
function on the right hand side of (\ref{eq:replace}).

Inserting the above results in eqs.(\ref{eq:ftdqcd}), (\ref{eq:g1dqcd}),
one obtains expressions for the perturbative contribution to the 
diffractive structure function. After 
carrying out the angular integrals in $\vec l_{t,n}$ and applying 
(\ref{eq:replace}), we find:
\begin{eqnarray}
F_T^D(\beta,Q^2,\xp,t)\bigg|_{t=0} & = &  \frac{Q^4}{xN_c}\; \sum_q e_q^2
\; \int \frac{\d k_t^2}{k_t^4}\; \alpha^2 (1-\alpha)^2 \; 
\frac{\alpha^2+(1-\alpha)^2}{|1-2\alpha|} \nonumber \\
& & \hspace{-2.2cm} \times \left[ \; \alpha_s\, \int \frac{\d l_t^2}{4l_t^2}
\left( \frac{k_t^2-\bar Q^2}{k_t^2+\bar Q^2} - \frac{k_t^2 - l_t^2-\bar Q^2}
{\sqrt{\left[\bar Q^2 + k_t^2 + l_t^2\right]^2 - 4 l_t^2 k_t^2}}\right)
 \; \phi_g(l_t^2,\xp) \right]^2\, ,\label{eq:ftdfull} \\
\beta g_1^D(\beta,Q^2,\xp,t)\bigg|_{t=0} & = & \frac{\beta Q^4}{4xN_c}\;
 \sum_q e_q^2 \; \int \frac{\d k_t^2}{k_t^4}\; \alpha^2 (1-\alpha)^2 \nonumber \\
& & \hspace{-4.5cm} \times
\Bigg[\;  -\alpha_s \, 2T_F\, \int \frac{\d l_{t,1}^2}{2l_{t,1}^2}
\; \Bigg\{ \, |1-2\alpha| \, \frac{l_{t,1}^2}{\bar Q^2} \left(
\frac{k_t^2-\bar Q^2}{k_t^2+\bar Q^2} - \frac{k_t^2 - l_{t,1}^2-\bar Q^2}
{\sqrt{\left[\bar Q^2 + k_t^2 + l_{t,1}^2\right]^2 - 4 l_{t,1}^2 k_t^2}}\right)
\nonumber \\
& & \hspace{-3.3cm}
+ \frac{\alpha^2+(1-\alpha)^2}{|1-2\alpha|}\, \frac{1}{\bar Q^2} \left(
\bar Q^2 + k_t^2 + l_{t,1}^2- \sqrt{\left[\bar Q^2 
+ k_t^2 + l_{t,1}^2\right]^2 
- 4 l_{t,1}^2 k_t^2}\right) \Bigg\} \phi_{\Delta g}(l_{t,1}^2,\xp)
\nonumber \\
& & \hspace{-4.1cm}
+ \alpha_s\, |1-2\alpha|\, C_F \,
\frac{ k_t^2}{\bar Q^2(\bar Q^2 + k_t^2)} 
\; \int \d l_{t,1}^2 \left( \phi_{\Delta q}(l_{t,1}^2,\xp) +
\phi_{\Delta \bar q}(l_{t,1}^2,\xp)\right) \Bigg]\nonumber \\
& & \hspace{-2.7cm} \times \left[\; \alpha_s\, \int \frac{\d l_{t,2}^2}
{4l_{t,2}^2}
\left( \frac{k_t^2-\bar Q^2}{k_t^2+\bar Q^2} - \frac{k_t^2 - l_{t,2}^2
-\bar Q^2}
{\sqrt{\left[\bar Q^2 + k_t^2 + l_{t,2}^2\right]^2 - 4 l_{t,2}^2 k_t^2}}\right)
 \; \phi_g(l_{t,2}^2,\xp) \right]\; , \label{eq:g1dfull}
\end{eqnarray}
where $\bar Q^2 = \alpha (1-\alpha) Q^2$. The parameter 
$\alpha$ is itself not an independent 
variable, but related to $k_t^2= \alpha(1-\alpha)Q^2(1-\beta)/\beta$. 
The result for $F_T^D$ is in agreement with earlier results in the 
literature~\cite{ftd,wusthoff}.

The integrations over the transverse momenta of the exchanged
($l_{t,i}$) and final state ($k_t$) partons extend into the region of 
small momenta, where the perturbative calculation is expected to be no
longer applicable. 
A simple model assumption for the infrared behaviour of the unintegrated 
structure functions $\phi(\l_t^2,\xp)$, as proposed for the unpolarized
diffractive structure function in~\cite{bekw}, allows however to 
extrapolate the 
expressions obtained above into the region of small $l_{t,i}$ and small 
$k_t$. We will demonstrate below that the resulting expression for 
$g_1^D$ turns out to be insensitive on the infrared behaviour of the 
polarized unintegrated distributions. 

Following the argumentation in~\cite{bekw}, we introduce a 
hadronic scale $k_0^2$, where the transition between soft and hard
dynamics takes place. Around this scale, the $l_t^2$-dependence of 
$\phi(l_t^2,\xp)$ changes from the perturbative $1/l_t^2$ behaviour 
(cf.~eq.~(\ref{eq:replace})) to a constant $1/k_0^2$:
\begin{equation}
\phi(l_t^2,\xp) \sim \frac{1}{k_0^2}\; \left( \frac{k_0^2}{l_t^2} 
\right)^{\nu(l_t^2/k_0^2)} \qquad \mbox{with}\; \nu(l_t^2/k_0^2)
\longrightarrow
 \left\{ \begin{array}{r@{\quad:\quad}l} 1 & l_t^2 \gg k_0^2 \\ 
0&  l_t^2 \ll k_0^2\end{array} \right.
\label{eq:phimodel}
\end{equation}
With this assumption, all $l_{t,i}^2$ integrals
remain finite, while preserving the 
leading logarithmic behaviour as given in
(\ref{eq:gllog})--(\ref{eq:dqbllog}). The 
 normalization of $\phi(l_t^2,\xp)$ 
in the infrared region is determined entirely by non-perturbative
effects and has
to be taken as free parameter.

Closer inspection of the $l_{t,i}^2$-integrands 
shows that for large $k_t^2$ the dominant contribution to the
unpolarized amplitude comes from the
region $k_0^2 < l_{t,2}^2 < k_t^2 + \bar Q^2$. 
The polarized gluon induced amplitude also changes its dependence on
$l_{t,i}^2$ around $l_{t,i}^2 = k_t^2 + \bar Q^2$. Unlike 
the unpolarized amplitude, it does however not acquire an additional
suppression factor $1/l_{t,i}^2$ for larger $l_{t,i}^2$, such that the 
relevant integration region for the polarized amplitudes is
bound only by the kinematical cut:
$k_0^2 < l_{t,1}^2 < \mbox{min}(\alpha,1-\alpha)\hat s$. 

These upper cuts on $l_{t,i}^2$
determine, at the leading logarithmic level, the scale
at which the target structure is probed. It is therefore appropriate 
to define the hard, perturbative contribution to the $k_t^2$-integral 
by demanding $k_t^2 + \bar Q^2 > k_0^2$. This perturbative contribution
to $F_T^D$ can be obtained by
retaining only the leading power in $l_{t,n}^2$ in the integrands. It
takes the well 
known~\cite{ftd,wusthoff} form
\begin{eqnarray}
F_{T,{\rm hard}}^D(\beta,Q^2,\xp,t)\bigg|_{t=0} & = & 
\frac{\beta}{\xp N_c}\, \sum_q e_q^2 \, \beta^2 (1-\beta) \,
\int_{k_0^2}^{\frac{Q^2}{4\beta}}
 \frac{\d \tilde k^2}{\tilde k^4} \; \frac{1-2\beta \tilde k^2/Q^2}
{\sqrt{1- 4\beta \tilde k^2/Q^2}}\nonumber\\
& & \times
\left[ \alpha_s \int_{k_0^2}^{\tilde k^2}
\d l_t^2 \, \phi_g (l_t^2,\xp) \right]^2\; , 
\label{eq:ftdll}
\end{eqnarray}
with $\tilde k^2 = k_t^2/(1-\beta)= k_t^2 + \bar Q^2$. 
The hard contribution to $g_1^D$ consists of two terms, corresponding to 
the regions with $l_{t,1}^2$ being smaller or larger than $\tilde k^2$:
\begin{equation}  
\beta g_{1,{\rm hard}}^D=
\beta g_{1,<}^D +  \beta g_{1,>}^D\; .
\end{equation}
These terms read:
\begin{eqnarray}
\beta g_{1,<}^D(\beta,Q^2,\xp,t)\bigg|_{t=0} & = &
 \frac{\beta}{4 \xp N_c}\, \sum_q e_q^2 \, \beta (1-\beta)\,
\int_{k_0^2}^{\frac{Q^2}{4\beta}} \frac{\d \tilde k^2}{\tilde k^4} 
\nonumber \\
& &  \times
\Bigg[ \alpha_s \int_{k_0^2}^{\tilde k^2}
\d l_{t,1}^2 \, \Bigg( -2T_F\, \frac{1-2\beta \tilde k^2/Q^2}
{\sqrt{1- 4\beta \tilde k^2/Q^2}}\;
\phi_{\Delta g}(\l_{t,1}^2,\xp)\nonumber \\
& & \hspace{0.7cm}
+ C_F\, \sqrt{1- 4\beta \tilde k^2/Q^2}\left(\phi_{\Delta_q}
(\l_{t,1}^2,\xp) + \phi_{\Delta_{\bar q}}(\l_{t,1}^2,\xp)\right) 
\Bigg) \Bigg] \nonumber \\
 & & \times \left[\alpha_s\; \int_{k_0^2}^{\tilde k^2} 
\d l_{t,2}^2 \phi_g (l_{t,2}^2,\xp) 
\right]\; ,
\label{eq:g1dll}\\
\beta  g_{1,>}^D(\beta,Q^2,\xp,t)\bigg|_{t=0} & = &
 \frac{\beta}{4 \xp N_c}\, \sum_q e_q^2 \, \beta (1-\beta)\,
\int_{k_0^2}^{\frac{Q^2}{4\beta}} \frac{\d \tilde k^2}{\tilde k^4} 
\, \sqrt{1- 4\beta \tilde k^2/Q^2}\nonumber \\
& & \hspace{-4.3cm} \times
\left[ \alpha_s \int_{\tilde k^2}^{\tilde k^2 /\xp}
\d l_{t,1}^2 \,
\left(-2T_F \, \phi_{\Delta g}(\l_{t,1}^2,\widehat{\xp})
+ C_F \left(\phi_{\Delta_q}
(\l_{t,1}^2,\widehat{\xp}) + \phi_{\Delta_{\bar q}}
(\l_{t,1}^2,\widehat{\xp})\right) 
\right) \right] \nonumber \\
& & \hspace {-4.3cm}
\times \left[\alpha_s\; \int_{k_0^2}^{\tilde k^2}
 \d l_{t,2}^2 \phi_g (l_{t,2}^2,\xp) 
\right].
\label{eq:g1dsl}
\end{eqnarray}

It is worth noting that $g_{1,>}^D$ corresponds to a region in phase
space that is usually discarded in calculations based on strong ordering 
in transverse momentum.
Similar terms are also present in 
the inclusive polarized structure function $g_1$ at 
small $x$~\cite{ber}, where they yield double logarithms of the form 
$\alpha_s^n \ln^{2n}(1/x)$. 
These logarithmic terms are potentially large at small $x$ (or small $\xp$ in
diffraction). A resummation of them, 
following closely the $g_1$ calculation of~\cite{ber}, 
will be outlined in section~\ref{sec:resum}.

Note in (\ref{eq:g1dsl}) the $\widehat{\xp}$ used as the 
argument of parton distributions $\phi_{\Delta j}$ ($j=g,q,\bar{q}$). 
This argument follows from the on-shell condition across the $t$-channel 
cut of the diffractive amplitude.  In the region of
$l_t^2\gg \tilde k^2$, an energy fraction $\xp + l_t^2/(\alpha \hat s)$ 
or $\xp + l_t^2/((1-\alpha) \hat s)$
is needed to fulfill this condition. To implement this energy shift 
consistently, we have to go back to (\ref{eq:dm2q}),
(\ref{eq:dm2qb}) and can no longer
sum up the terms  corresponding to the quark-proton centre-of-mass energy
$s_{qp}=\alpha \hat{s}$ and  $s_{qp}=(1-\alpha)\hat{s}$.
Instead of $\phi_{\Delta_q}(l^2_t,\xp)+\phi_{\Delta_{\bar q}}(l^2_t,\xp)$ 
and $\phi_{\Delta_g}(l^2_t,\xp)$,
one finds 
\begin{eqnarray}
\phi_{\Delta_q}(l^2_t,\widehat{\xp})+\phi_{\Delta_{\bar q}}
(l^2_t,\widehat{\xp}) &= &
\frac{1}{2} \Bigg( 
      \phi_{\Delta_q}\left(l^2_t,\xp+\frac{l_t^2}{\alpha_-\hat s}\right)
    + \phi_{\Delta_q}\left(l^2_t,\xp+\frac{l_t^2}{\alpha_+\hat s}\right)
\nonumber \\
&& \hspace{0.25cm}
    + \phi_{\Delta_{\bar q}}\left(l^2_t,\xp+\frac{l_t^2}{\alpha_-\hat s}\right)
    + \phi_{\Delta_{\bar q}}\left(l^2_t,\xp+\frac{l_t^2}{\alpha_+\hat s}\right)
  \Bigg) \nonumber \\
&&\hspace{-2.52cm}
   + \frac{1}{2 \sqrt{1- 4\beta \tilde k^2/Q^2}} \Bigg(
    - \phi_{\Delta_q}\left(l^2_t,\xp+\frac{l_t^2}{\alpha_-\hat s}\right)
    + \phi_{\Delta_q}\left(l^2_t,\xp+\frac{l_t^2}{\alpha_+\hat s}\right)
\nonumber \\
&& \hspace{1.2cm}
    - \phi_{\Delta_{\bar q}}\left(l^2_t,\xp+\frac{l_t^2}{\alpha_-\hat s}\right)
    + \phi_{\Delta_{\bar q}}\left(l^2_t,\xp+\frac{l_t^2}{\alpha_+\hat s}\right)
\Bigg)\; ,\nonumber\\
\phi_{\Delta_g}(l^2_t,\widehat{\xp}) & = &
\frac{1}{2}\left( 
      \phi_{\Delta_g}\left(l^2_t,\xp+\frac{l_t^2}{\alpha_-\hat s}\right)
    + \phi_{\Delta_g}\left(l^2_t,\xp+\frac{l_t^2}{\alpha_+\hat s}\right)
   \right)\;,
\label{eq:kincorr}
\end{eqnarray}
where $\alpha_{\pm} = 1/2(1\pm \sqrt{1- 4\beta \tilde k^2/Q^2})$. 
This effect will be taken into account in the next section 
when we discuss the resummation of the double logarithms.

Up to now, we have discussed the hard, perturbative contribution
to the diffractive structure functions. These can be probed only by 
restricting the diffractive final state such that a minimum cut on
$\tilde k^2$ is realized, for example by demanding a pair of high transverse
momentum jets. The inclusive structure function also receives
contributions from the soft, non-perturbative region $\tilde k^2 < k_0^2$.
These can not be calculated from first principles; using the 
model assumption (\ref{eq:phimodel}) for the infrared behaviour of the
unintegrated structure functions, it is however possible to extend 
(\ref{eq:ftdfull}) and (\ref{eq:g1dfull}) into the soft region. 
This procedure will enable us to 
 estimate the $\beta$-dependence of 
the soft contribution to $g_1^D$, the absolute normalization of this 
contribution is however determined entirely by non-perturbative effects
and cannot be calculated within our approach. 

To facilitate the discussion of the soft contributions, let us rewrite
(\ref{eq:ftdfull}) and (\ref{eq:g1dfull}) by introducing 
amplitude functions $\psi_i(\alpha,k_t^2,l_t^2)$:
\begin{eqnarray}
F_T^D(\beta,Q^2,\xp,t)\bigg|_{t=0} & = & \frac{1}{\xp N_c} \sum_q \, e_q^2
\; \beta \; \int \frac{\d
  \tilde k^2}{(1-\beta)} \nonumber\\ & & \times
\left[\alpha_s\; \int \frac{\d l_t^2}{l_t^2} 
\psi_g(\alpha,k_t^2,l_t^2)
 \phi_g(l_t^2,\xp) \right]^2 \; ,\label{eq:ftdampl}\\
\beta 
g_1^D(\beta,Q^2,\xp,t)\bigg|_{t=0} & = & \frac{1}{4\xp N_c} \sum_q \, e_q^2
\; \beta^2 \; \int \frac{\d
 \tilde k^2}{(1-\beta)} 
\nonumber \\ & & \times 
 \Bigg[\alpha_s\; \int \frac{\d
  l_{t,1}^2}{l_{t,1}^2} 
\Bigg\{\psi_{\Delta g}(\alpha,k_t^2,l_{t,1}^2)
 \phi_{\Delta g}(l_{t,1}^2,\xp) \nonumber \\
& & \hspace{0.4cm} + \psi_{\Delta q}(\alpha,k_t^2,l_{t,1}^2)
 \phi_{\Delta q}(l_{t,1}^2,\xp) + \psi_{\Delta \bar q}(\alpha,k_t^2,l_{t,1}^2)
 \phi_{\Delta \bar q}(l_{t,1}^2,\xp) \Bigg\} 
\Bigg]\nonumber \\
&  &  \times
\left[\alpha_s\; \int \frac{\d l_{t,2}^2}{l_{t,2}^2} 
\psi_g(\alpha,k_t^2,l_{t,2}^2)
 \phi_g(l_{t,i}^2,\xp) \right] \; .\label{eq:g1dampl}
\end{eqnarray}

We already discussed the behaviour of the amplitude functions 
in the context of the 
perturbative contributions to $g_1^D$ and $F_T^D$. For small $\tilde
k^2$, they become:
\begin{eqnarray}
\psi_g(\alpha,k_t^2,l_t^2) & = & 
 \left\{ \begin{array}{r@{\quad:\quad}l} \beta (1-\beta)
     \; \l_t^2/\tilde k^2  & l_t^2 \ll \tilde k^2 \; ,\\ 
(1-\beta)/2 &  l_t^2 \gg \tilde k^2\end{array} \right.\\
\psi_{\Delta g}(\alpha,k_t^2,l_t^2) & = & -\frac{2T_F}{C_F} \; 
 \psi_{\Delta q,\bar q}(\alpha,k_t^2,l_t^2)\nonumber \\
&=  & -2T_F\frac{1-\beta}{\beta}\; \frac{l_t^2}{\tilde k^2}\quad:\quad
\mbox{all} \quad l_t^2 \; .
\end{eqnarray}
These simple forms allow us to identify the dominant regions 
in the $l_{t,i}^2$ and $\tilde k^2$ integrations by mere power counting.

In the region of 
$\tilde k^2< k_0^2$, 
the dominant contribution to the $l_{t,i}^2$-integral for the
unpolarized amplitude comes from $\tilde k^2 < l_{t,i}^2 < k_0^2$:
\begin{equation}
\int_{\tilde k^2}^{k_0^2} \frac{\d l_{t,i}^2}{l_{t,i}^2} 
\psi_g(\alpha,k_t^2,l_{t,i}^2)
 \phi_g(l_{t,i}^2,\xp)  \sim (1-\beta) \frac{1}{k_0^2}\;  \ln
 \frac{k_0^2}{\tilde k^2} \; .
\label{eq:unpolsoft}
\end{equation}
For the polarized amplitude one finds that, even for $\tilde k^2 <
k_0^2$,
 the $l_{t,i}^2$-integral is dominated by 
the perturbative region $k_0^2 < l_{t,i}^2 < \tilde k^2/\xp$. 
The non-perturbative behaviour of $\phi_{\Delta g, \Delta q,\Delta \bar
  q}$ becomes relevant only if $\tilde k^2 < \xp k_0^2$. 
The contribution 
from this latter region is however suppressed by a factor $1/\ln^2
\xp$ with respect to the contribution from the former region. 
The soft contribution to $\beta g_1^D$ is thus given simply by
extrapolating (\ref{eq:g1dsl}) into the region $\xp k_0^2< \tilde k^2 <
k_0^2$.   

However, some caution is to be taken, since this procedure extrapolates 
the amplitudes into a region of phase space corresponding to transverse
momenta of the order $\Lambda_{{\rm QCD}}$. Although the polarized
amplitudes are (in contrast to the unpolarized amplitude) well behaved
in this region, they may still be changed 
due to parton motion or confinement. 
Such effects can easily modify the right hand side of
(\ref{eq:unpolsoft}), where $k_0^2$ and $\tilde k^2$ 
could be accompanied by terms of order $\Lambda_{{\rm QCD}}$, thus
imposing a natural limit on the accuracy of the soft interpolation. 

\section{Resummation of $\ln^2 (1/\xp)$}
\label{sec:resum}
Perturbative QCD corrections to the polarized inclusive structure function 
$g_1(x,Q^2)$ at small $x$ contain leading double logarithmic 
terms of the form 
$\alpha_s^n \ln^{2n}(1/x)$. 
In spacelike diffractive processes, such terms can only 
appear in polarized structure functions; the most singular terms in 
the unpolarized singlet case take the form $\alpha_s^n \ln^{n}(1/x)$.
The $\alpha_s^n \ln^{2n}(1/x)$ terms
arise from regions in phase space which are 
discarded  in the
conventional strong ordering in transverse momentum.
A resummation of the leading double logarithmic terms in 
$g_1(x,Q^2)$ at small $x$ has been performed in~\cite{ber}, using an
infrared evolution equation~\cite{kl} for the polarized exchange
amplitudes.

From the spin dependence of the target trace (\ref{eq:ptrace}), one
would expect similar exchange amplitudes also to appear in $g_1^D$. 
In fact, the occurrence of terms from non-strongly ordered regions in 
(\ref{eq:g1dsl}) is a first manifestation of
a double  logarithmic enhancement already at the leading order in
$\alpha_s$. In the following, we shall briefly outline how these terms
can be resummed to all orders. To facilitate the discussion, we shall
assume that the polarized distributions for all quark flavours are
identical at small $x$, i.e.~we will only consider the singlet quark
distribution. A resummation of non-singlet contributions would follow 
exactly the same procedure as outlined below.  

Let us consider the combination of unintegrated polarized
distributions appearing in (\ref{eq:g1dsl}),\footnote{Note 
that the notation used here differs
  from~\protect\cite{ber}. The {\it vectors}
 $T$ and $R$ used there are obtained from our {\it matrices} $T$ and $R$ 
 by contraction with $(2e_q^2,0)$. Furthermore, all evolution matrices are
 transposed compared to~\cite{ber}.}
\begin{eqnarray}
\lefteqn{   \, \sum_q e_q^2 \,
 \int^{\mu^2} \d l_t^2  \left[ -2 T_F \, \phi_{\Delta g}(\l_{t}^2,\xp)
+ C_F \left( \phi_{\Delta_q}
(\l_{t}^2,\xp) + \phi_{\Delta_{\bar q}}(\l_{t}^2,\xp)\right)\right]}  
\nonumber \\
& = &  \sum_q \frac{e_q^2}{n_f} \left( 1 ,  0 \right)
\left( \begin{array}{cc} C_F & -2T_F n_f  \\ 2 C_F & 4C_A \end{array} \right)
\int^{\mu^2} \d l_t^2 \left( \begin{array}{c} \phi_{\Delta \Sigma}
(\xp, l_t^2) \\
\phi_{\Delta g} (\xp,l_t^2) \end{array} \right) \nonumber\\
& \equiv &   \sum_q \frac{e_q^2}{n_f}
 \left( 1 ,  0 \right) T^{LO} (\xp,\mu^2,\mu_0^2)
\int^{\mu_0^2} \d l_t^2 \left( \begin{array}{c} 
\phi_{\Delta \Sigma}(\xp, l_t^2) \\
\phi_{\Delta g} (\xp,l_t^2) \end{array} \right) \; ,
\label{eq:comb1}
\end{eqnarray}
where $n_f$ is the number of active flavours and $\Delta \Sigma = \sum_q 
(\Delta q + \Delta \bar q$). 
The above equation 
coincides with the contribution from the first rung of an ordinary DGLAP
ladder amplitude (evolved from given boundary conditions at 
$\mu_0^2$ to larger $\mu^2$)
to the inclusive structure function $g_1(x,Q^2)$ in the small-$x$ 
limit~\cite{ross}. 
In addition to this term, (\ref{eq:g1dll}) also contains
\begin{eqnarray}
\lefteqn{-\sum_q e_q^2 \; 2 T_F\; \int^{\mu^2}
 \d l_t^2 \phi_{\Delta g}(\l_{t}^2,\xp)}
\nonumber \\
 &=&  \sum_q \frac{e_q^2}{n_f} \; \frac{T_Fn_f}{2(T_Fn_f+C_A)} \; (2, -1) \;
 T^{LO} (\xp,\mu^2,\mu_0^2)
\int^{\mu_0^2} \d l_t^2 \left( \begin{array}{c} 
\phi_{\Delta \Sigma}(\xp, l_t^2) \\
\phi_{\Delta g} (\xp,l_t^2) \end{array} \right) \; .
\end{eqnarray}
Resummation of the double leading logarithmic contributions turns
$T^{LO}$ into a resummed evolution matrix $T$.

This evolution matrix
can be expressed by its Mellin transformation
$R(\omega,\mu^2,\mu_0^2)$ as
\begin{equation}
T(\xp,\mu^2,\mu_0^2) = \int_{-i\infty}^{+i\infty}
\frac{\d \omega}{2\pi i}\, \left(\frac{s_{qp}}{\mu^2}
\right)^\omega\, \frac{e^{-i\pi\omega}-1}{2} \, R
(\omega,\mu^2,\mu_0^2)\; .
\label{eq:mellin}
\end{equation}
The matrix $R$ obeys an
infrared  evolution equation, given by (3.7)
in~\cite{ber}:
\begin{equation}
\left(\omega - \frac{\partial}{\partial \ln \mu^2} \right)
\, R = \frac{1}{8\pi^2} F_0 R\; ,
\label{eq:iee}
\end{equation}
where the evolution kernel matrix $F_0$ contains all partonic 
colour singlet $t$-channel exchanges at small $x$. 
$F_0$ is itself determined by an 
evolution equation for four parton amplitudes at small $x$, as
calculated in~\cite{ber}. 

The matrix $F_0$ can be diagonalized to
\begin{equation}
\hat F_0 = {\rm diag} (\lambda_0^+, \lambda_0^-)\qquad \mbox{with} \qquad
F_0 = E_0^{-1} \hat F_0 E_0\,
\end{equation}
where $E_0$ is the matrix of eigenvectors of $F_0$. Using these, we can
solve the evolution equation (\ref{eq:iee}):
\begin{equation}
R(\omega,\mu^2,\mu_0^2) = E_0^{-1} \, \frac{1}{\omega - \hat F_0/(8\pi^2)}
\, \left(\frac{\mu^2}{\mu_0^2}\right)^{\hat F_0/(8\pi)} E_0
\left( \begin{array}{c} 2e^2_q\\0\end{array} \right)\; ;
\end{equation}
transformation according to (\ref{eq:mellin}) then yields the resummed 
evolution matrix $T$. Some numerical studies on $R$ can be found
in~\cite{ber}. 

The resummed expression for $g_1^D$ is obtained by introducing
the vectors 
\begin{equation}
T^{(a)}  =  (1,0) T\; , \qquad
T^{(b)} = \frac{T_Fn_f}{2(T_Fn_f+C_A)} (2,-1) T \; ,
\end{equation}
and replacing the integrals over the unintegrated structure functions 
in (\ref{eq:g1dll}) and (\ref{eq:g1dsl})  by the
corresponding resummed expressions. 
$T^{(a)}$ and $T^{(b)}$ 
should be considered as the sum of two terms corresponding to the 
quark energies  $s_{qp}=\alpha \hat{s}$ and  
$s_{qp}=(1-\alpha)\hat{s}$, as listed in (\ref{eq:kincorr}).

Recall that in the double logarithmic 
approximation  the negative signature amplitude sums up 
not only 
 the ladder graphs but also  nonladder 
 diagrams with  transverse momenta of the embraced
  gluons $q_t^2>\mu^2$ 
as well~\cite{kl}. For $\mu^2>k_t^2$ all such contributions are 
already included into the amplitude $T$ (\ref{eq:mellin}). The presence
of the spectator quark in the diffractive amplitude could in principle
yield additional nonladder contributions. In these contributions, the
large gluon momentum $q_t$ has  however to appear in two quark
propagators, such that these terms do not bear a leading logarithm.
Thus the presence of the 
spectator quark does not destroy the structure of the double logarithmic 
amplitudes.

The perturbative contributions from
different regions in $l_{t,1}^2$ can moreover be combined into a
single expression. The resulting double leading logarithmic expressions
read: 
\begin{eqnarray}
\beta g_{1,{\rm hard},{\rm DLL}}^D(\beta,Q^2,\xp,t) \bigg|_{t=0} & = & 
 \frac{\beta}{4 \xp N_c}\, \sum_q e_q^2 \, \beta (1-\beta)\,
\int_{k_0^2}^{\frac{Q^2}{4\beta}} \frac{\d \tilde k^2}{\tilde k^4} 
\, \sqrt{1- 4\beta \tilde k^2/Q^2}\nonumber \\
& & \hspace{-5.5cm} \times
\left[ \alpha_s \left( T^{(a)} (\xp, \tilde k^2/\xp, k_0^2)
+ \frac{2\beta \tilde k^2/Q^2}{1-4\beta \tilde k^2/Q^2} 
T^{(b)} (\xp, \tilde k^2, k_0^2) \right) \left( 
\begin{array}{c} \Delta \Sigma (\xp, k_0^2) \\
\Delta g (\xp, k_0^2)\end{array} \right) \right]\nonumber \\
& & \hspace{-5.5cm}
 \times \left[\alpha_s\; \int_{k_0^2}^{\tilde k^2}
 \d l_{t,2}^2 \phi_g (l_{t,2}^2,\xp) \right]\; ,\\
\beta g_{1,{\rm soft},{\rm DLL}}^D(\beta,Q^2,\xp,t) \bigg|_{t=0} & = & 
 \frac{\beta}{8 \xp N_c}\, \sum_q e_q^2 \, (1-\beta)\,
\int_{\xp k_0^2}^{k_0^2} \frac{\d \tilde k^2}{\tilde k^2} \nonumber \\
& & \hspace{-3cm} \times
\left[ \alpha_s \; T^{(a)} (\xp, \tilde k^2/\xp, k_0^2)
\left( 
\begin{array}{c} \Delta \Sigma (\xp, k_0^2) \\
\Delta g (\xp, k_0^2)\end{array} \right)\right] \nonumber \\
& & \hspace{-3cm}
 \times \left[\alpha_s\; \int_{\tilde k^2}^{k_0^2}
 \frac{\d l_{t,2}^2}{l_{t,2}^2} \phi_g (l_{t,2}^2,\xp) \right] \; .
\end{eqnarray}
It is evident from the above expressions, that contributions containing
$T^{(b)}$ appear only 
for large $\tilde k^2$, which is realized e.g.~in diffractive jet
production. For most other observables, such as for example the
diffractive structure function itself, or diffractive vector meson
production~\cite{teubner,ryskin}, 
this term plays only a minor role. In these latter cases, the
resummed diffractive 
amplitude becomes directly proportional to the inclusive polarized 
structure function at small $x$.

\section{Conclusions and Outlook}

In this paper we have calculated the longitudinal spin-spin asymmetry for 
the heavy 
photon to $q\bar{q}$-pair diffractive dissociation in the framework of a 
perturbative two parton exchange model.
The spin asymmetry is given by 
the interference between unpolarized and polarized diffractive amplitudes.
The spin dependent amplitude has a double logarithmic form and includes the 
contribution from the region of inverse $k_t$ ordering 
where the transverse momentum of the $t$-channel parton $l_t$ is larger than 
the transverse momentum $k_t$ of the outgoing quark. We provide explicit 
expressions for strongly ordered and inversely ordered contributions to
the diffractive structure function $g_1^D$ in terms of unintegrated 
structure functions. $g_1^D$ receives contributions from both polarized 
quark and gluon distributions at small $x$. Like in the inclusive
structure function $g_1$, a positive $\Delta g$ at small $x$ results in
a negative diffractive spin asymmetry.

We present an expression which provides the full summation of all double 
logarithmic contributions in the spin dependent amplitude at small $x$. 
The expression can smoothly be continued into the infrared region and 
matched with the soft part of the amplitude.

Our results show that the diffractive asymmetry increases towards small 
$\beta$: $A^D\sim 1/\beta$.
We must however emphasize that in this paper only the $\gamma^*\to q\bar{q}$ 
dissociation is considered. To avoid contributions from more 
complicated final states our results for the
asymmetry should be used only in the  large-$\beta$ domain ($\beta
>0.3$, i.e.\  $M^2 > 2Q^2$) where the contamination of $q\bar{q}\, +\, g$ 
states is presumably small.

A numerical estimate of the asymmetry can be obtained from
eq.~(\ref{eq:ftdll}) and (\ref{eq:g1dll}) by approximating the
unintegrated parton distributions at leading $\ln Q^2$,
(\ref{eq:gllog})--(\ref{eq:dqbllog}). Using recent parametrizations of
unpolarized and polarized parton distributions, one finds a resulting
asymmetry $A_L$ of the order $-10^{-2}\ldots -10^{-3}$ 
at $Q^2\sim 10$ GeV$^2$ and $\xp\sim 10^{-3}$, which is at least
an order of magnitude above the non-perturbative estimates
of~\cite{mana}. The estimated asymmetry is moreover rather insensitive
on the precise value used for the matching scale $k_0^2$. It must
however be kept in mind that this estimate relies on the behaviour of
the polarized gluon distribution at small $x$, which is not known 
directly 
from experiment at present, but can be at best inferred indirectly from
the observed evolution of the polarized structure function. 
Like in the inclusive structure function $g_1$~\cite{kwie}, the 
resummation of  $\ln^2(1/\xp)$ will result in a substantial enhancement
 of the asymmetry.

\section*{Acknowledgements}
The work of MGR was supported by the Russian Fund of Fundamental 
Research (98 02 17629), the INTAS grant  
95 - 311, by the Volkswagen-Stiftung and by DESY. 
JB was supported in part by the EU Fourth Framework Programme
`Training and Mobility of Researchers', Network `Quantum Chromodynamics and
the Deep Structure of Elementary Particles', contract FMRX-CT98-0194 (DG 12 -
MIHT).

\end{document}